\documentclass[%
 aip,
 amsmath,amssymb,
 reprint,%
]{revtex4-1}

\usepackage{graphicx}% Include figure files
\usepackage{dcolumn}% Align table columns on decimal point
\usepackage{bm}% bold math
\usepackage{amsmath}
\usepackage{amssymb}
\usepackage{graphicx}
\usepackage{epstopdf}
\usepackage{suffix}
\usepackage{mathtools}
\usepackage{gensymb}
\usepackage[utf8]{inputenc}
\usepackage[T1]{fontenc}
\usepackage{mathptmx}

\begin{document}

\preprint{AIP/123-QED}

\title{Magnonics in collinear magnetic insulating systems}
% Force line breaks with \\

\author{B. Flebus}
 \affiliation{Department of Physics, Boston College, 140 Commonwealth Avenue Chestnut Hill, MA 02467}%Lines break automatically or can be forced with \\

\date{\today}% It is always \today, today,

\begin{abstract}
In the last decades, collinear magnetic insulating systems have emerged as promising energy-saving information carriers.  Their elementary collective spin excitations, i.e., magnons, can propagate for long distances  bypassing the Joule heating effects that arise from electron scattering in metal-based devices. 
This tutorial article provides an introduction to  theoretical and experimental advances in the study of magnonics in collinear magnetic insulating systems. We start by outlining the quantum theory of spin waves in ferromagnetic and antiferromagnetic systems and we discuss their quantum statistics. We review the phenomenology of spin and heat transport of the coupled coherent and incoherent spin dynamics and  the interplay between magnetic excitations and lattice degrees of freedom. Finally, we introduce the reader to the key ingredients of two experimental probes of magnetization dynamics, spin transport  and NV-center relaxometry setups, and discuss experimental findings relevant to the outlined theory.
\end{abstract}

\maketitle

\section{\label{sec:level1}Introduction}

The possibility of using the electron's spin as a new degree of freedom for transmitting information has received significant attention in the last decades~[\onlinecite{Spin},\onlinecite{Spin1}]. Of particular interest for novel device applications are magnetic insulating systems~[\onlinecite{insulator},\onlinecite{magnonics}]: solids in which electrons are “frozen” in their atomic positions while a pure spin current, i.e., a flow of spin angular momentum, can be transmitted through a wave-like collective motion of the electrons’ spins. Transmitting signals via spin flows in insulators has the potential to circumvent the heating and energy loss that is generated when the electrons collide, e.g., with impurities, as they move within a crystal lattice. 
The discovery of the spin Hall effect (SHE) and the inverse spin Hall effect (ISHE)~[\onlinecite{SHEref,SHEref1,SHEref2,SHEref3,SHEref4,SH1,SH2}], which convert a charge current into a spin current and \textit{vice versa}, has further advanced the field by allowing  electrical injection and detection of spin currents in heterostructures comprised of magnetically ordered insulators interfaced with a normal metal. More recently, Uchida and co-authors showed that a pure spin flow  can be generated in a magnetic insulator not only via electrical injection, but by heat gradients as well via the spin Seebeck effect (SSE)~[\onlinecite{SSE}], suggesting the possibility of converting waste heat into spin signals and, thus, opening up intriguing prospects for “greener” information technology~[\onlinecite{reviewCal}]. \\
Experimental observations of electrically- and thermally-generated spin transport have confirmed the potential of magnetic insulating systems as long-range information carriers~[\onlinecite{spindiff1,spindiff2,spindiff3,AFdifflength}].

There is a plethora of magnetic insulating systems, which display ground-state spin configurations ranging from collinear to non-collinear depending on the intrinsic spin-spin interactions. Non-collinear configurations stem from spin couplings that can endow the system with a non-trivial topological band structure and engender exotic spin phases and textures. However,  for the purpose of spin current transmission, these systems are not ideal platforms due to the lack of spin conservation; thus, the magnetic materials currently deployed in spin and heat transport setups provide  spins that are mainly collinear.

In this tutorial we aim at introducing the readers to the key properties of collective spin excitations in collinear ferromagnetic and antiferromagnetic  insulating systems. In section I, we outline the fundamental ingredients of the theory of spin waves  and we review the coupled equations describing the coherent (long-wavelength) magnetization dynamics and the spin and heat transport driven by the thermal magnon cloud. Finally, we briefly discuss  the coupling between magnetic and lattice degrees of freedom.

In section II, we introduce spin transport and NV relaxometry setups as probes of magnetization dynamics and we discuss recent experimental results making use of the theoretical tools outlined in section I.

\section{\label{sec:level2} Collinear magnetic insulators}

In this section, we introduce the fundamental principles of spin-wave theory taking as examples  monoatomic ferromagnetic and  two-sublattice antiferromagnetic collinear insulators. We discuss the interaction between their coherent and incoherent spin dynamics, and  their coupled spin and heat transport properties. Finally, we review the coupling between magnetic and lattice degrees of freedom. 

\subsection{$U$(1)-symmetric ferromagnetic systems}

We consider  a $U$(1)-symmetric ferromagnetic  insulating system. At each  $i$th site of the monoatomic lattice, we can define a spin  $\mathbf{S}_{i}=S \mathbf{n}_{i}$,   where $S$ is the classical spin  and the unit vector $\mathbf{n}_{i}$  the spin orientation. At a temperature $T$ far below the magnetic ordering (Curie) temperature $T_{c}$, we can map spin fluctuations around the ground state onto second-quantized operators via the Holstein Primakoff (HP) transformation~[\onlinecite{Holstein}]. For a ground state with spins uniformly oriented along the $-z$-direction, the HP transformation reads, in the macrospin limit, as
\begin{align}
\hat{S}_{i z}=\hat{a}_{i}^{\dagger} \hat{a}_{i}-S, \; \; \; \; \; \; \; \;    \hat{S}_{i-}=\frac{\hat{S}_{ix}+i\hat{S}_{iy}}{2} \simeq \sqrt{2S} \hat{a}_{i}\,,
\label{HP}
\end{align}
where  $\hat{a}_{i}^{\dagger}$ ($\hat{a}_{i}$) is the canonical bosonic second-quantized operator that creates (annihilates) a magnon carrying spin angular momentum $\hbar$. In the limit of small excitation amplitude, i.e., for $T \ll T_{c}$, the magnon-magnon interaction
corrections are small~[\onlinecite{magnoninteraction}]. Thus, any term higher than quadratic in the magnon creation and annihilation operators can be omitted from the magnetic Hamiltonian $\mathcal{H}_{\text{m}}$, i.e.,
\begin{align}
\mathcal{H}_{\text{m}}=\sum_{\mathbf{k}} \hbar \omega_{\mathbf{k}} \hat{a}^{\dagger}_{\mathbf{k}} \hat{a}_{\mathbf{k}}+E_{0}\,.
\end{align}
Here $E_{0}$ is the ground state classical energy,   $\omega_{\mathbf{k}}$ the spin-wave dispersion, and 
\begin{align}
\hat{a}_{\mathbf{k}}=\frac{1}{\sqrt{N}} \sum_{i} e^{i \mathbf{k} \cdot \mathbf{r}_{i}} \hat{a}_{i}, \; \; \; \; \;   \; \; \hat{a}^{\dagger}_{\mathbf{k}}=\frac{1}{\sqrt{N}} \sum_{i} e^{-i \mathbf{k} \cdot \mathbf{r}_{i}} \hat{a}^{\dagger}_{i}\,,
\end{align}
where $N$ is the number of lattice sites.

The  Hamiltonian $\mathcal{H}_{\text{m}}$ possesses $U$(1) symmetry, i.e.,  $[\hat{S}_{iz}, \mathcal{H}_{\text{m}}]=0$, i.e., it is invariant for rotations in spin space around the $z$-axis. Owing to this symmetry, Noether's theorem guarantees that the $z$-component of the spin, $S_{iz}=\langle \hat{S}_{iz} \rangle $, and, consequently, the total magnon number,  $\tilde{N}=\sum_{\mathbf{k}}\langle \hat{a}_{\mathbf{k}}^{\dagger} \hat{a}_{\mathbf{k}} \rangle $, are conserved quantities, where $\langle ... \rangle$ stands for the equilibrium (thermal) average.   Such conservation law is, however,  an approximation:  in reality,  magnon-magnon and magnon-lattice interactions  that invalidate spin conservation are present in any insulating system. Even in the long-wavelength and low-temperature limit, one must include a dimensionless Gilbert damping  parameter $\alpha$, with $\alpha \ll 1$, in the Landau-Lifshitz-Gilbert equation~[\onlinecite{LLG1},\onlinecite{LLG2}] in order to account for the experimentally observed broadening of the ferromagnetic resonance (FMR), i.e.,
\begin{align}
\dot{\mathbf{n}}(\mathbf{r})=-\gamma \mathbf{n}(\mathbf{r}) \times \mathbf{H}(\mathbf{r})- \alpha  \mathbf{n}(\mathbf{r}) \times \dot{\mathbf{n}}(\mathbf{r})\,.
\label{LLG}
\end{align}
Here, $\gamma$(>0) is  the gyromagnetic ratio, $ \mathbf{H}(\mathbf{r})$ is the effective (Landau-Lifshitz) field  acting on the  order parameter $\mathbf{n}(\mathbf{r})$, with $\mathbf{n}(\mathbf{r})=\mathbf{n}_{i}$ in the continuum limit. 
Equation~(\ref{LLG}) can be solved in the linear regime, i.e., by neglecting terms  quadratic in spin fluctuations, i.e., $n_{i} n_{j}$ for $i,j=x,y$.  By performing a HP and Fourier transformation~(\ref{HP}), one can  find a direct correspondence between the linearized classical dynamics~(\ref{LLG}) and the second-quantized Hamiltonian via the Heisenberg relation
\begin{align}
\dot{\hat{a}}_{\mathbf{k}}=\frac{i}{\hbar} \left[ \mathcal{H}_{\text{m}}, \hat{a}_{\mathbf{k}} \right]\,.
\label{Heisenberg}
\end{align}
According to Eqs.~(\ref{LLG}) and (\ref{Heisenberg}), however, non-Hermitian terms violating magnon number conservation will appear in the Hamiltonian $\mathcal{H}_{\text{m}}$~[\onlinecite{FlebusNH}]. Until recently, such terms have been neglected in the quantum theory of spin waves; a discussion of such non-Hermicity will be addressed elsewhere.

 Here  we focus on magnetic systems with low damping,  for which spin nonconserving terms can be (approximately) neglected. Namely, we discuss the limit in which the energy scales of dissipative processes are very small compared to the exchange interactions that control the thermalization of the magnon distribution function. It has been extensively debated whereas the elementary quanta of spin waves, i.e., magnons, can be described as quasi-particles whose density is (approximately) conserved in collinear magnetic systems with low damping. The assumption of magnon density (quasi-) conservation carries important consequences. From a statistical standpoint, it implies the existence of a well-defined magnon chemical potential, and thus  of magnon Bose-Einstein condensation, achievable even at relatively high temperatures~[\onlinecite{BEC}]. Magnon density conservation has also been predicted to underlie a spin superfluid phase that shares many similarities with supercurrent of electric charge in superconductors and the mass superflow in helium~[\onlinecite{SpinSup,SpinSup1,Yuan}].

Recent experiments have shown that the injection of spin current into a collinear magnet can drive the magnon gas into a quasi-equilibrium state described by a Bose–Einstein statistics with non-zero chemical potential~[\onlinecite{Cornelissen2016,chemDem,Du2017}], suggesting that the inclusion of a chemical potential is necessary to properly capture the experimental features of  incoherent magnon transport. 
Thus, we will describe thermal magnons as a (quasi-equilibrium) thermalized Bose-Einstein ensemble with a well-defined chemical potential $\mu$, i.e.,
 \begin{align}
 n_{\text{BE}}\left( \frac{\hbar \omega_{\mathbf{k}}-\mu}{k_{B}T} \right)\,,
 \end{align}
 where  $n_{\text{BE}}(x)=\left(e^{x}-1\right)^{-1}$ is the Bose-Einstein distribution function and  $k_{B}$ the Boltzmann constant.
 The equation governing the diffusive dynamics of thermal magnons density $\tilde{n}$ can be written as
\begin{align}
\dot{ \tilde{n}}(\mathbf{r}) + \boldsymbol{\nabla} \cdot \mathbf{j}_{s}(\mathbf{r})=-g_{n \mu} \mu(\mathbf{r})  \,.
\label{diff}
\end{align}
Here,  $g_{n \mu}$ parametrizes the spin relaxation rate and  $\mathbf{j}_{s}(\mathbf{r})=-\sigma \boldsymbol{\nabla} \mu(\mathbf{r})\,$  the spin current, where  $\sigma$ is the magnon spin conductivity. 
Thermal and long-wavelength magnons interact, e.g., via exchange coupling or single-ion anisotropies. 
The form of such interaction between coherent (\ref{LLG}) and incoherent (\ref{diff}) spin dynamics  can be derived phenomenologically, using the symmetry and reciprocity principles discussed in Ref.~[\onlinecite{Flebus2016}]. For a $U$(1)-symmetric system, it suffices to supplement Eqs.~(\ref{LLG}) and~(\ref{diff}) with the following terms [\onlinecite{Flebus2016}]
\begin{align}
\hbar \dot{\mathbf{n}}(\mathbf{r})=&-\eta \mathbf{n}(\mathbf{r}) \times \left[ \hbar \dot{\mathbf{n}}(\mathbf{r})-\mu(\mathbf{r}) \mathbf{z} \times \mathbf{n}(\mathbf{r}) \right]\,, \label{132} \\
\dot{\tilde{n}}(\mathbf{r})=&-\eta \tilde{s} \mathbf{z} \cdot \mathbf{n}(\mathbf{r}) \times \left[ \dot{\mathbf{n}}(\mathbf{r}) -\frac{\mu(\mathbf{r})}{\hbar} \mathbf{n}(\mathbf{r}) \times \mathbf{z} \right]\,,
\label{133}
\end{align}
where $\eta \ll 1$ is a phenomenological (dimensionless) coefficient parametrizing the strength of interactions between the order parameter $\mathbf{n}$ and the thermal magnon cloud. Here, we have introduced the reduced spin density $\tilde{s}=S/V-\tilde{n}$, with $V$ being the system volume.
Equations~(\ref{132}) and~(\ref{133}) show that the magnon chemical potential can be tuned by driving the coherent spin dynamics. The later, in turn, can pump spin angular momentum into the incoherent thermal magnon cloud~[\onlinecite{Flebus2016}]. Using this mechanism, Du and coauthors measured the magnon chemical potential via single spin magnetometry based on nitrogen-vacancies~[\onlinecite{Du2017}], as we will describe in detail later.

\subsection{$U$(1)-symmetric antiferromagnetic systems}

Due to their lack of a net magnetic moment, antiferromagnets can not be controlled via an external magnetic field with the same ease as ferromagnetic systems. However,  the absence of production of stray fields,  which  limit the packing density of ferromagnetic elements, and their fast spin dynamics ($\sim$THz) make them  desirable for the development of spintronic devices~[\onlinecite{reviewAF},\onlinecite{reviewAF1}].  To illustrate the fundamental properties of antiferromagnetic magnons, here we focus on  a $U$(1)-symmetric antiferromagnetic insulator with a ground-state \textit{staggered} magnetic order. The neighbouring spins are aligned in opposite directions: this magnetic arrangement can be conveniently described in terms of a unit cell with
two magnetic sublattices, $A$ and $B$. For each $i$th unit cell, we can define the spin $\mathbf{S}_{A(B)_{i}}= S \mathbf{n}_{A(B)i}$, where the unit vector $\mathbf{n}_{A(B)i}$ represents the spin orientation of the magnetic sublattice $A(B)$. We consider sublattices with the same magnitude of magnetization;  if the sublattices have different magnetizations, i.e., $|\mathbf{S}_{A}|\neq |\mathbf{S}_{B}|$, the system is ferrimagnetic 
 and in several respects behaves like a ferromagnet,  e.g., yttrium iron garnet (YIG)~[\onlinecite{YIG}].
 
\begin{figure}[t!]
\includegraphics[width=1\linewidth]{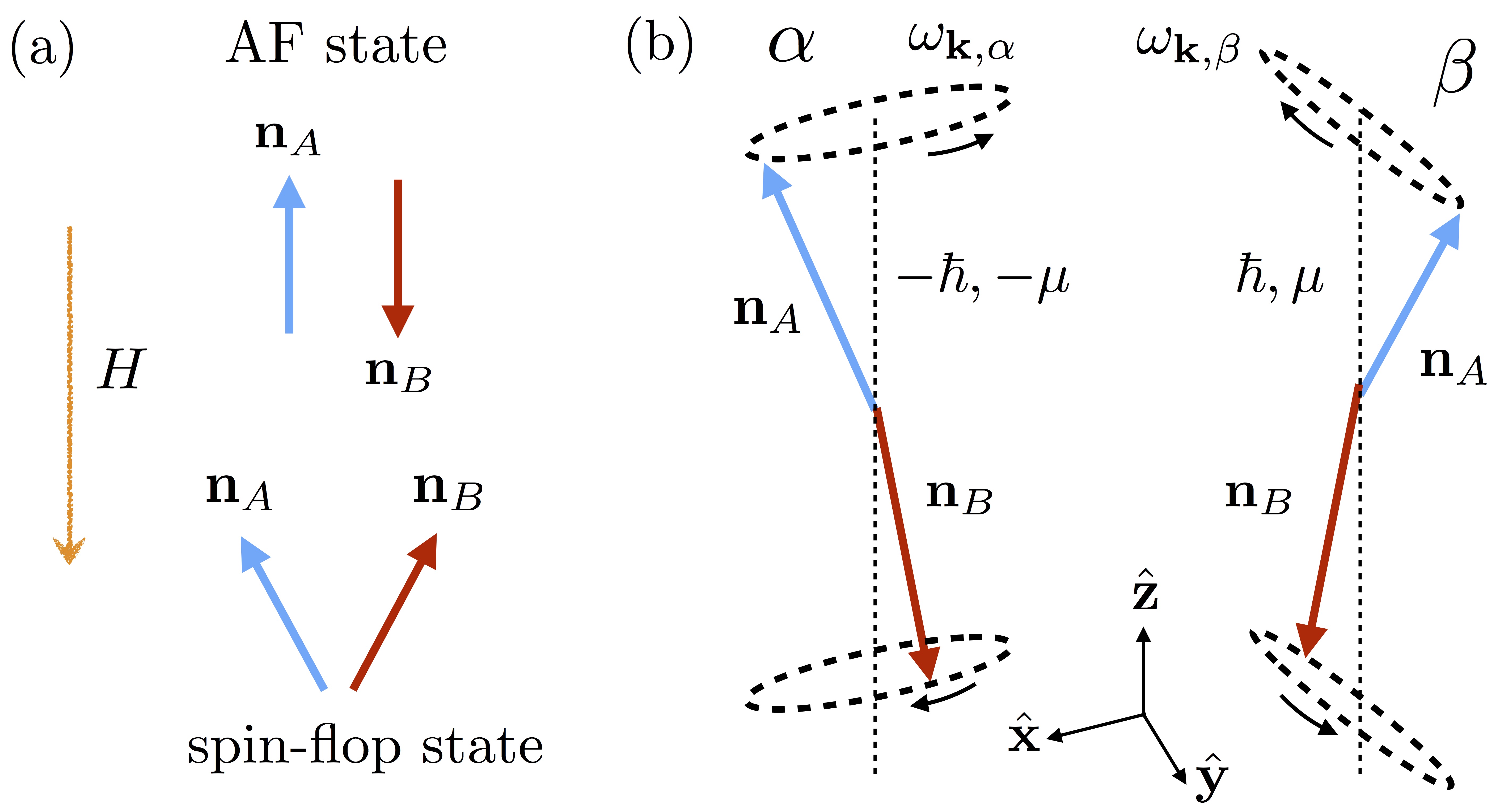}
\caption{(a) Spin configurations in collinear antiferromagnets subjected to a magnetic field $H$ oriented along the magnet symmetry axis. Below the spin-flop transition, the order parameters $\mathbf{n}_{A}$ and $\mathbf{n}_{B}$ are antiparallel. Increasing the magnetic field induces a spin-flop transition, leading to a canted spin structure. (b)  Magnon eigenmodes, labelled as $\alpha$ and $\beta$. The antiferromagnetic mode $\alpha$ ($\beta$), which carries $\pm \hbar$ spin angular momentum, can be visualized as a combination of both up and down spins, precessing counterclockwise (clockwise) at the frequency $\omega_{\alpha (\beta)}$. The two antiferromagnetic modes are associated with an equal and opposite chemical potential $\mu$.}
\label{Fig1}
\end{figure}

We  set the orientations of the order 
parameters as $\mathbf{n}_{A(B)i} \parallel \pm \hat{\mathbf{z}}$. This assumption is valid only below the critical field at which a spin-flop transition occurs, i.e., when the two-sublattice spins rotate suddenly to a direction perpendicular to the easy-magnetization direction, as shown in Fig.~\ref{Fig1}(a).
Below the spin-flop transition and far below the magnetic ordering (N\'eel) temperature $T_{N}$,  
 the HP transformation reads, in the macrospin limit, as  
\begin{align}
\hat{S}_{Aiz}&= S- \hat{a}_{i}^{\dagger} \hat{a}_{i}, \; \; \; \; \; \; \; \;   \hat{S}_{A+}=\sqrt{2S} \hat{a}_{i}\,,\nonumber \\
\hat{S}_{Biz}&= \hat{b}_{i}^{\dagger} \hat{b}_{i}-S, \; \; \; \; \; \; \; \;   \hat{S}_{B+}=\sqrt{2S} \hat{b}_{i}^{\dagger}\,,
\label{HPAF}
\end{align}
where $\hat{a}_{i}^{\dagger}$ ($\hat{a}_{i}$) and $\hat{b}^{\dagger}$ ($\hat{b}$) are the creation (annihilation) 
operators for spin deviations on the sublattice $A$ and $B$, respectively.  The sublattice spins are coupled by, e.g., nearest-neighbor exchange interactions; thus,  terms of the form $\hat{a}_{\mathbf{k}} \hat{b}_{-\mathbf{k}}$ ($\hat{a}_{\mathbf{k}}^{\dagger} \hat{b}_{-\mathbf{k}}^{\dagger}$) appear in the Hamiltonian  upon HP and Fourier transformation. The magnetic Hamiltonian  can be brought in a diagonal form, i.e., 
\begin{align}
\mathcal{H}_{\text{m}}=\sum_{\mathbf{k}} \left[ \hbar \omega_{\mathbf{k},\alpha} \hat{\alpha}^{\dagger}_{\mathbf{k}} \hat{\alpha}_{\mathbf{k}}+ \hbar \omega_{\mathbf{k},\beta} \hat{\beta}^{\dagger}_{\mathbf{k}} \hat{\beta}_{\mathbf{k}} \right] \,,
\end{align} 

via a Bogoliubov-de-Gennes transformation~[\onlinecite{BdG}], i.e.,
\begin{align}
\hat{\alpha}_{\mathbf{k}}=u_{\mathbf{k}} \hat{a}_{\mathbf{k}}-v_{\mathbf{k}} \hat{b}^{\dagger}_{-\mathbf{k}}\,, \; \; \; \; \; \; \;  \; \; \; \; 
\hat{\beta}_{\mathbf{k}}=u_{\mathbf{k}} \hat{b}_{\mathbf{k}}-v_{\mathbf{k}} \hat{a}^{\dagger}_{-\mathbf{k}}\,,
\label{BdG}
\end{align}
where $\hat{\alpha}_{\mathbf{k}}^{\dagger}$ ($\hat{\beta}_{\mathbf{k}}^{\dagger}$) is the creation 
operator for a  magnon mode with dispersion $\omega_{\mathbf{k}, \alpha(\beta) }$.  
Setting  $u^2_{\mathbf{k}}-v^2_{\mathbf{k}}=1$ guarantees  the Bogoliubov-de-Gennes operators to obey the canonical commutation relations.
By rewriting the $z$-component
of the spin  operator (in units of $\hbar$)  in terms of the Bogoliubov quasi-particles (\ref{BdG}) as~[\onlinecite{Rezende}]
\begin{align}
\hat{S}_{z}=\sum_{i} \left( S_{Aiz} + S_{Biz} \right)=\sum_{\mathbf{k}} \hbar \left( -\hat{\alpha}^{\dagger}_{\mathbf{k}} \hat{\alpha}_{\mathbf{k}}+\hat{\beta}^{\dagger}_{\mathbf{k}} \hat{\beta}_{\mathbf{k}} \right)\,,
\label{AFsz}
\end{align}
we can identify bosonic operator $\hat{\alpha}_{\mathbf{k}}^{\dagger}$ $(\hat{\beta}_{\mathbf{k}}^{\dagger})$ as the operator creating a magnon with spin angular momentum $\mp \hbar$. $U$(1) symmetry implies the conservation of the $z$-component of the total spin  $S_{z}=\langle \hat{S}_{z} \rangle$~(\ref{AFsz}).
Generally, the magnon chemical potential accounts for how much the free energy of the system in thermal equilibrium will change by adding a quanta of spin angular momentum. In an antiferromagnetic system, net spin injection is achieved by creating an imbalance between the population of the two magnon species~(\ref{AFsz});  thus, there is a single chemical potential associated with both magnonic species~[\onlinecite{FlebusCP}]. 
 Assuming (approximate) spin conservation, we can assign a chemical potential $\mp \mu$ to the Bose-Einstein distribution function $n_{\text{BE}, \alpha (\beta)}$ of the antiferromagnetic eigenmode $\alpha$ ($\beta$)~[\onlinecite{FlebusCP}], i.e.,
\begin{align}
n_{\text{BE},\alpha(\beta)}\left( \frac{\hbar \omega_{\alpha (\beta) k}\pm \mu}{k_{B}T}\right)\,.
\end{align} 

 Thermal fluctuations engender a finite thermal magnon density of both antiferromagnetic modes, i.e., $\tilde{n}_\alpha$ and $\tilde{n}_\beta$; since the contributions of the two modes to the spin current have opposite signs, it is convenient to rewrite the spin diffusion equation, in the continuum limit, in terms of the net density of thermal magnons carrying angular momentum $\hbar$, i.e., $\tilde{n}=\tilde{n}_{\beta}-\tilde{n}_{\alpha}$, as
\begin{align}
\dot{ \tilde{n}}(\mathbf{r}) + \boldsymbol{\nabla} \cdot \mathbf{j}_{s}(\mathbf{r})=-g_{\mu} \mu(\mathbf{r})  \,,
\label{diffAF}
\end{align}
which is formally analogous to Eq.~(\ref{diff}).  Their classical coupled dynamics  can be modeled by the phenomenological Landau-Lifshitz equation~[\onlinecite{kittelAF}]
\begin{align}
\dot{\mathbf{n}}_{A(B)}=-\gamma \times \left[ \mathbf{H}_{A(B)}-H_{c} \mathbf{n}_{B(A)} \right]\,,
\label{LLGAFF}
\end{align}
where $\mathbf{H}_{A(B)}$ is the effective (Landau-Lifshitz) field acting on the order parameter $\mathbf{n}_{A(B)}$ and $H_{c}$ is the, e.g., exchange, field coupling the spin dynamics of the two sublattices.

The linearized dynamics of the long-wavelength magnetic order parameters~(\ref{LLGAFF}) can be diagonalized in terms of two normal modes $\mathbf{n}_{\alpha}$ and $\mathbf{n}_{\beta}$ ~[\onlinecite{kittelAF}].  In the mode $\alpha$ both up and down spins precess clockwise with frequency $\omega_{\alpha \mathbf{k}}$, while in the $\beta$ mode the spins undergo a counterclockwise precession with frequency $\omega_{\beta \mathbf{k}}$, as depicted in Fig.~\ref{Fig1}(b).  The corresponding Landau-Lifshitz-Gilbert equation reads as ~[\onlinecite{kittelAF},\onlinecite{RezendeAFF}] 
\begin{align}
\dot{\mathbf{n}}_{\alpha (\beta)}=-\gamma \mathbf{n}_{\alpha (\beta)} \times \mathbf{H}- \alpha_{\alpha, \beta}  \mathbf{n}_{\alpha (\beta)} \times \dot{\mathbf{n}}_{\alpha (\beta)}\,,
\label{LLGAF}
\end{align}
where we have included a mode-selective Gilbert damping parameter $\alpha_{\alpha(\beta)}$~[\onlinecite{GilbertAF}]. 
 In terms of the normal modes, the interplay between coherent and incoherent spin dynamics can be accounted for by supplementing Eqs.~(\ref{diffAF}) and ~(\ref{LLGAF})  with~[\onlinecite{FlebusCP}]
\begin{align}
\hbar \dot{\mathbf{n}}_{\alpha}=&- \eta_{\alpha} \mathbf{n}_{\alpha} \times ( \hbar \dot{\mathbf{n}}_{\alpha}-\mu \mathbf{z} \times \mathbf{n}_{\alpha} ) \,,\label{191} \\
\hbar \dot{\mathbf{n}}_{\beta}=&- \eta_{\beta} \mathbf{n}_{\beta} \times ( \hbar \dot{\mathbf{n}}_{\beta}-\mu \mathbf{z} \times \mathbf{n}_{\beta} ) \,,   \label{192} \\  
\dot{\tilde{n}}
=&-\eta_{\alpha} \tilde{s}_{\alpha} \mathbf{z} \cdot \mathbf{n}_{\alpha} \times ( \dot{\mathbf{n}}_{\alpha}- \frac{\mu}{\hbar} \mathbf{n}_{\alpha}\times \mathbf{z}) \nonumber \\
& -\eta_{\beta} \tilde{s}_{\beta} \mathbf{z} \cdot \mathbf{n}_{\beta} \times ( \dot{\mathbf{n}}_{\beta}- \frac{\mu}{\hbar} \mathbf{n}_{\beta}\times \mathbf{z})\,, \label{193}
\end{align}
with $\tilde{s}_{\alpha (\beta)}=s - \tilde{n}_{\alpha (\beta)}$. Here, $\eta_{\alpha(\beta)} \ll 1$ is a phenomenological (dimensionless) coefficient parametrizing the strength of interactions between the order parameter $\mathbf{n}_{\alpha (\beta)}$ and the thermal magnon cloud.

\subsection{Coupled spin and heat transport}

\begin{figure}[t!]
\includegraphics[width=1\linewidth]{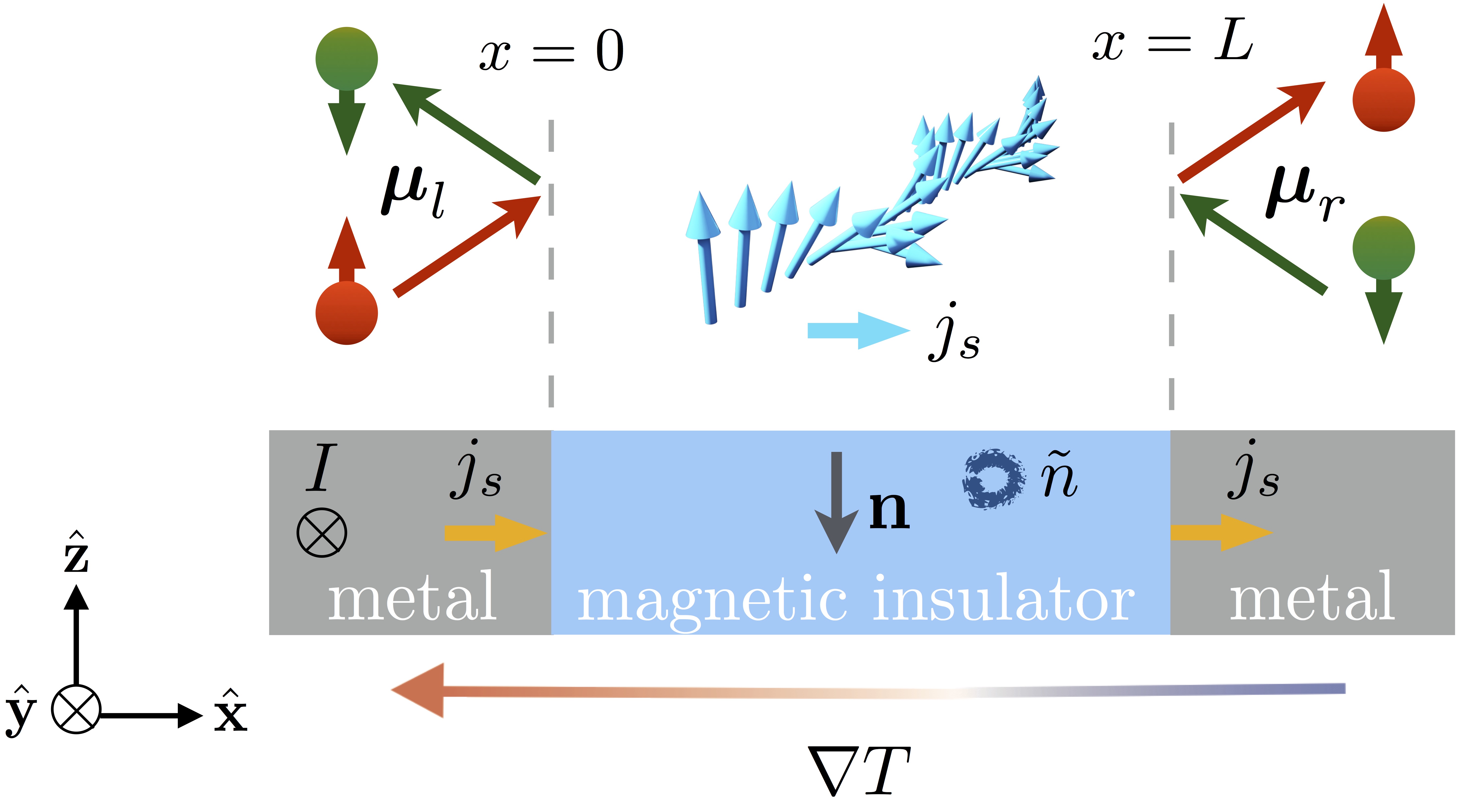}
\caption{ Metal|magnetic insulator|metal heterostructure.  Via the spin Hall effect, a charge current $I$ sent through the metal along the $y$-direction generates a spin accumulation $\boldsymbol{\mu_{l}}=\mu_{l} \hat{\mathbf{z}}$ at the metal$|$magnetic insulator interface, which injects a $z$-polarized spin current $j_{s}$  into the magnetic insulator. Due to the Joule heating associated with the charge current,  a temperature gradient $\nabla T$ is set across the sample. The injected spin current and the temperature gradient trigger the dynamics of the magnetic order parameter $\mathbf{n}$  and of the thermal magnon cloud density $\tilde{n}$. The coherent and incoherent spin dynamics at the magnetic insulator$|$metal interface pump a spin current $j_{s}$ at $x=L$, leading to a spin accumulation $\boldsymbol{\mu}_{r}=\mu_{r} \hat{\mathbf{z}}$ that can be converted into a measurable voltage in the metal via the inverse spin Hall effect.}
\label{FigTransport}
\end{figure}

Spin and heat transport in collinear magnetic insulators have been extensively investigated both theoretically and experimentally in a variety of setups. Here, for simplicity,  we present the theory of coupled spin and heat transport by considering a normal metal$|$magnetic insulator$|$normal metal heterostructure in a 1D geometry, shown in Fig.~\ref{FigTransport}. The metallic reservoirs act as thermal baths, set at two different temperatures, i.e.,  $T_{l}$ and $T_{r}$. The temperature bias  generates a linear temperature gradient $\nabla T $ across the sample. The  normal metals have strong spin–orbit coupling:  if a charge current flows through  them,  a spin accumulation is generated at the metal$|$magnetic insulator interface and \textit{vice versa}. 

The bulk spin, $\mathbf{j}_{s}$, and heat, $\mathbf{j}_{q}$, currents carried by magnons in the magnetic insulating system can be written as 
\begin{align}
\begin{pmatrix}\mathbf{j}_{s} \\ \mathbf{j}_{q} \end{pmatrix} = - \begin{pmatrix} \boldsymbol{\sigma} & \boldsymbol{\zeta} \\ \boldsymbol{\rho}&\boldsymbol{\kappa} \end{pmatrix} \begin{pmatrix} \boldsymbol{\nabla} \mu \\ \boldsymbol{\nabla} T \end{pmatrix}\,,
\label{203}
\end{align}
where the tensors $\boldsymbol{\sigma}$, $\boldsymbol{\kappa}$ , $\boldsymbol{\zeta}$, and $\boldsymbol{\rho}$($= T\boldsymbol{\zeta}$ by the Onsager-Kelvin 
relation~[\onlinecite{Onsager}, \onlinecite{Callen}]) are, respectively, the spin and (magnetic) heat conductivities, and the spin Seebeck and Peltier coefficients.  
The hydrodynamics equations for spin and heat transport can be easily constructed within the Boltzmann transport theory~[\onlinecite{Boltzmann}], i.e.,
\begin{align}
\dot{\tilde{n}}+\boldsymbol{\nabla} \cdot \mathbf{j}_{s}=-g_{n\mu} \mu - g_{nT} \left( T - T_{p} \right)\,, \nonumber \\
\dot{u}+\boldsymbol{\nabla} \cdot \mathbf{j}_{q}=-g_{u \mu} \mu - g_{uT} \left( T - T_{p} \right)\,.
\label{210}
\end{align}
Here, $T$ ($T_p$) is the magnon (phonon) temperature, $u$  the energy density of the thermal cloud, while $g_{n (u) \mu }$ and $g_{n (u) T}$  parametrize, respectively,  the relaxation of magnons via inelastic magnon-magnon and magnon-phonon interactions  and the magnon thermalization  to the phonon temperature. 
The relations~(\ref{203}) and (\ref{210}), complemented by Eqs.~(\ref{LLG},\ref{132},\ref{133})  and by Eqs.~(\ref{LLGAF},\ref{191},\ref{192},\ref{193}) for, respectively, a ferromagnetic and an antiferromagnetic system,   describe the coupled bulk spin and heat 
transport of the coherent and incoherent spin dynamics. 

The transport equations must be determined consistently with the boundary conditions for spin and heat transport at the 
interfaces. 
For a ferromagnet, we take the order parameter to be aligned along the $z$-direction, i.e., $\mathbf{n} \parallel - \hat{\mathbf{z}}$. For a two-sublattice antiferromagnet, it is convenient to 
introduce the N\'eel unit vector, i.e., $\mathbf{n}=\left( \mathbf{n}_{A}-\mathbf{n}_{B} \right)/2$. In this geometry,  the relevant nonequilibrium  spin accumulation at the metal$|$magnetic insulator interface at $x=0 (L)$ is polarized along $z$-direction~[\onlinecite{spinpumping},\onlinecite{spinpumping1}], i.e., $\boldsymbol{\mu}_{l(r)}=\mu_{l(r)} \hat{\mathbf{z}}$.  The  boundary conditions read as 
\begin{align}
 j_{s}|_{x=0 (L)} =& G_{l(r)} \left( \mu_{l(r)}-\mu \right)|_{x=0 (L)} + S_{l(r)} \left( T_{l(r)}-T \right)|_{x=0 (L)}\,, \nonumber \\
 j_{q}|_{x=0(L)} =& K_{l(r)} \left(T_{l(r)}-T\right)|_{x=0 (L)} +\Pi_{l(r)} \left( \mu_{l(r)}- \mu \right)|_{x=0 (L)}\,.
 \label{238}
\end{align}
Here, the coefficients  $G_{l(r)}$, $K_{l(r)}$, $S_{l(r)}$, and $\Pi_{l(r)}$(=$TS_{l(r)}$ by Onsager reciprocity) are the  magnon spin and  thermal conductances and spin Seebeck and Peltier coefficients, respectively, at the metal$|$magnetic insulator interface at $x=0(L)$.   Furthermore, the coherent magnetization dynamics pumps a spin current into an adjacent conductors~[\onlinecite{SpinSup1},\onlinecite{spinpumping,spinpumping1,spinpumping2,spinpumpingAF2}], i.e., 
\begin{align}
j_{s}|_{x=0(L)}=-g_{l(r)} \hat{\mathbf{z}} \cdot \left( \mathbf{n} \times \dot{\mathbf{n}} \right)|_{x=0 (L)}\,,
\end{align}
where $g_{l(r)}$ is the spin pumping efficiency of the metal$|$magnetic insulator interface at $x=0(L)$,  related to the magnon spin conductance $G_{l(r)}$ by Onsager reciprocity~[\onlinecite{spinpumping1},~\onlinecite{spinpumping2}]. The bulk and interfacial transport coefficients appearing, respectively, in Eqs.~(\ref{203}) and ~(\ref{238}) can be obtained via algebraic manipulation and integration of the magnon Boltzmann equation over momentum. Their estimate depends on the details of the transport regime under investigation; for a detailed discussion of some examples, we refer the reader to Refs.~[\onlinecite{Cornelissen2016}] and~[\onlinecite{Flebus2016}]. 

It is worth remarking that the outlined phenomenology can be easily extended to a setup comprised of an insulating non-magnetic substrate$|$magnetic insulator$|$normal metal, commonly used, e.g., in longitudinal SSE measurements. At the substrate|magnetic insulator interface, the spin flow is blocked as there are no spin carriers in the substrate. Nevertheless, heat still can be transmitted via inelastic spin-preserving scattering processes between magnons and phonons. The corresponding boundary conditions at the substrate$|$insulator interface at $x=0$ read as
\begin{align}
j_{s}|_{x=0}=0\,, \; \; \; \; \; \; \;  \; \; \; j_{q}|_{x=0} =& K_{l} \left(T_{l}-T\right)|_{x=0}\,.
\end{align}

\subsection{Magnon-phonon coupling}

Coupling between collective magnetic and elastic excitations, i.e., magnons and phonons, is ubiquitous in magnetic insulating systems. It relies on relativistic effects such as dipole-dipole interactions and spin-orbit coupling, as well as on the dependence of the spin-spin exchange interactions on the phonon coordinates, and it provides a  pathway for thermalization and dissipation of the magnetization dynamics. While the exact form  of the phonon-driven magnon thermalization and dissipation rates  can be computed from microscopic models~[\onlinecite{phononmagnonlifetime}], often they are accounted for by introducing phenomenological parameters, see, e.g., Eq.~(\ref{210}). 

However, magnon thermalization and dissipation are not the only relevant mechanisms emerging from the magnetoelastic coupling.
When the magnetoelastic coupling term  that is quadratic in magnon and phonon operators does not vanish,  magnons and phonons can hybridize to form quasiparticles that are an admixture thereof, dubbed as magnon-polarons~[\onlinecite{mp1}, \onlinecite{mp2}]. The Hamiltonian of a coupled magnon-phonon system reads generally as
\begin{align}
\mathcal{H}=\mathcal{H}_{\text{m}}+\mathcal{H}_{\text{ph}}+\mathcal{H}_{\text{mec}}\,,
\label{MEC}
\end{align}
where $\mathcal{H}_{\text{m} (\text{ph})}$ is the magnon (phonon) quadratic Hamiltonian and $\mathcal{H}_{\text{mec}}$ the long-wavelength magnetoelastic coupling Hamiltonian, whose explicit form depends on the symmetries of the underlying lattice. The Hamiltonian~(\ref{MEC}) can be brought into diagonal form
via a Bogoliubov-de-Gennes transformation~[\onlinecite{BdG}], which maps  magnon and phonon operators into composite quasi-particles, i.e., magnon-polarons. 
\begin{figure}[t!]
\includegraphics[width=1\linewidth]{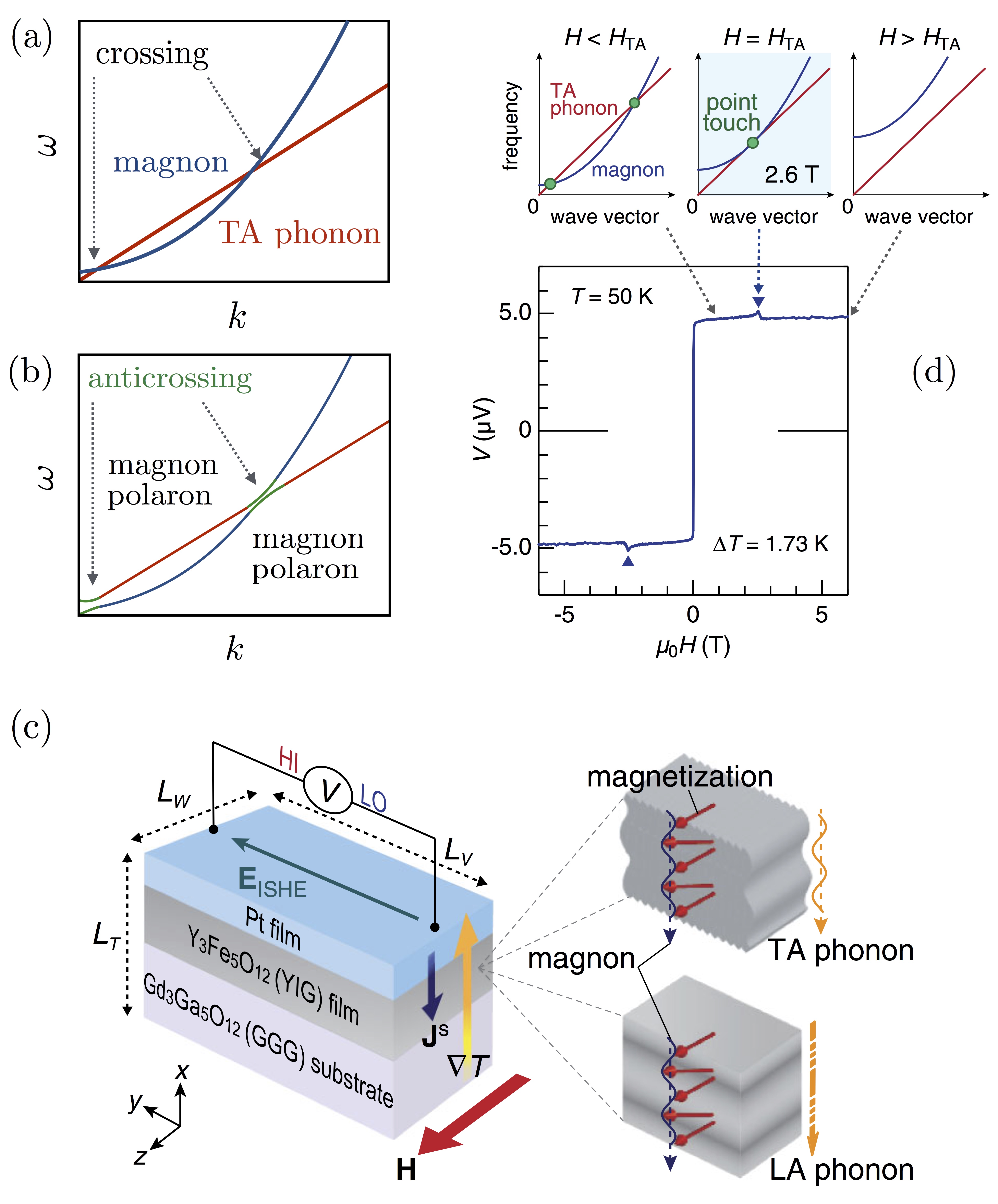}
\caption{(a)  The dispersions of uncoupled magnons and transverse acoustic (TA) phonons are degenerate at certain wavevectors. (b) In the presence of magnon-phonon coupling, the system can be diagonalized in terms of magnon-polarons, i.e., hybridized magnon-phonon quasi-particles. The magnetoelastic coupling lifts the degeneracy between the magnonic and phononic bands and the crossing points become anticrossing points.  At the anticrossing points, the mixing between magnons and phonons is maximized, while at other wave-vectors the magnon-polaron bands retain a phonon- or magnon-like character. (c) Longitudinal spin Seebeck setup comprised of a magnetic insulating system (YIG) sandwiched between   a gadolinium gallium garnet  (GGG) substrate and a platinum (Pt) film. A temperature gradient set across the sample generates a spin current that can be detected as a voltage in the Pt film~[\onlinecite{MP1}]. (d) At the magnetic field $H_{\text{TA}}$, the magnon dispersion shifts upwards such that the TA phonon branch becomes tangential and the effects of the magnetoelastic coupling are maximized, i.e., the magnon and phonon modes become strongly coupled over a relatively large volume in momentum space. The voltage measured in a spin Seebeck setup displays a peak at the field  $H_{\text{TA}}$: the magnon-polaron spin transport properties are enhanced with respect to their purely magnonic counterpart~[\onlinecite{MP1}]. }
\label{FigMP}
\end{figure} 
In the absence of coupling, transversal acoustic (TA) and longitudinal acoustic (LA) phonon bands are degenerate with the spin-wave dispersion at certain energies, as depicted in Fig.~\ref{FigMP}(a). When the coupling is sufficiently strong, the mutual interaction between the magnons and phonons lifts the degeneracy, leading to an anticrossing around which the magnon-polaron bands deviate from the purely phononic and magnonic ones, as shown in Fig.~\ref{FigMP}(b).
\begin{figure*}[htbp]
    \centering
    \includegraphics[width=0.9\linewidth]{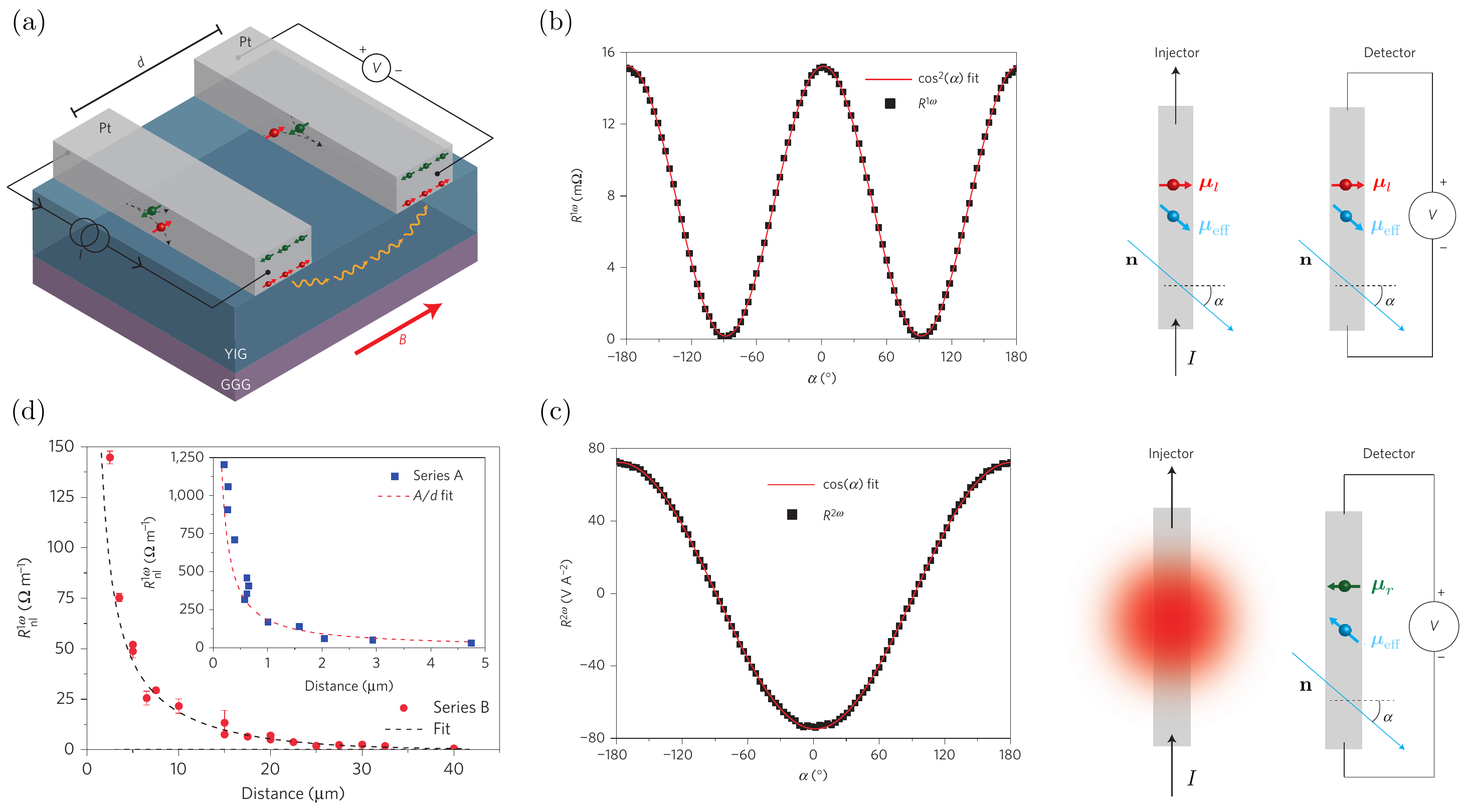}
    \caption{(a) Non-local spin transport setup and experimental data~[\onlinecite{spindiff1}]. Two Pt strips, separated by a distance $d$, are deposited on top of a magnetic insulating film (YIG) grown on a GGG substrate. One metallic strip (injector) injects spin current into the magnetic insulating system. The spin current propagates through the insulator and generates a voltage $V$ in the other metallic strip (detector). (b,c) By tuning the orientation of the in-plane magnetic field $\mathbf{B}$, the magnetic insulator order parameter $\mathbf{n}$ can be oriented at a generic angle $\alpha$ within the sample plane, i.e., $\mathbf{n} \parallel \cos \alpha$. A charge current in the injector leads to a spin accumulation $\boldsymbol{\mu}_{l}=\mu_{l} \hat{\mathbf{z}}$, with $\mu_{l} \propto I$, at the metal$|$magnetic insulator interface; the effective spin accumulation that injects non-equilibrium magnons in the insulator is parallel to the order parameter, i.e., $\boldsymbol{\mu}_{\text{eff}}=\mu_{l} \cos \alpha$. Similarly, a spin accumulation parallel to $\mathbf{n}$ is generated at the detector. The charge current $I$ yields as well  a temperature gradient $\nabla T \propto I^2$ that injects non-equilibrium magnons with no dependence on the angle $\alpha$. (b) The  first harmonic signal, which measures electrically-generated spin waves, is a product of the effects at the injector and detector, leading to a $\cos^2 \alpha$ dependence. (c) The measured second harmonic signal, which measures thermally-generated spin waves, displays a $\cos \alpha$ dependence. (d) The measured first harmonic signal is fitted to the model in Eq.~(\ref{FIT}).  }
    \label{SpinTransport}
\end{figure*}
Generally, strong hybridization occurs only in proximity of these anticrossings points and it is not likely to affect relevantly spin and heat transport coefficients~(\ref{210}), whose evaluation requires integration of the Boltzmann transport equation over the full reciprocal space.  However, it was recently found that, at critical magnetic field values, the magnon and phonon dispersions tangentially touch each other and the magnon–phonon hybridization effects are maximized~[\onlinecite{MP1}, \onlinecite{MP2}]. In correspondence of these fields, spin transport properties can be enhanced or suppressed, depending on the ratio between the phonon and the magnon relaxation times. If the phonon relaxation time is longer (shorter) than the magnon relaxation time, magnon-polarons will have a longer (shorter) lifetime than the pure magnetic excitations  and thus spin transport properties will be enhanced (suppressed).  These features have been recently observed in the field dependence of thermally generated current in YIG in a longitudinal SSE transport setup, shown in Figs.~\ref{FigMP}(c) and (d). Similar effects have been investigated as well  in antiferromagnetic systems~[\onlinecite{MPAF,MPAF1,MPAF2}].

%Various references for writing the coupling depending on the lattice structure ~(). 
%
%However, depending on the spin and acoustic energy dispersion and the energy scale one focuses on, more exotic effects can appear. 
%We note that this description crossing-anti-crossing etc is very general and can be applied to magnons photon etc. 
%Ferromagnetic and antiferromagnetic systems on square lattice. More recently, hexagonal systems. 
%Transport signatures. \\
%SeKwon mention of how the magneto-elastic coupling does interesting things topologically. \textbf{Let us read a second}
%
%Berry curvature can be induced in the anticrossing regions of magnon and phonon bands~\onlinecite{Berryphonons}

\section{Probing spin dynamics}

In this section we review two experimental probes of magnetization dynamics. We introduce the reader to spin transport setups, by focusing on the measurements of the YIG spin diffusion length~[\onlinecite{spindiff1}] and of thermally-generated spin currents in the antiferromagnetic insulator $\text{MnFe}_{2}$~[\onlinecite{AFdifflength}].  Secondly, we discuss a new minimally-invasive magnon sensing technique, i.e., NV-center relaxometry,  and review the recent measurements of the magnon chemical potential in YIG~[\onlinecite{Du2017}] and of the spin diffusion length in antiferromagnetic $\alpha$-$\text{Fe}_2\text{O}_3$~[\onlinecite{WangAF}].

\subsection{Spin transport}

A non-local transport set-up is commonly used as a probe of spin transport properties of magnetic insulating systems. This setup is comprised of two metallic strips, operating, respectively, as spin current injector and detector. The metals  are deposited on top of a magnetic insulating sample grown on a substrate, as shown in Fig.~\ref{SpinTransport}(a). A charge current $I$ modulated at frequency $\omega$ is sent through the injector. The SHE converts the charge current into a  spin accumulation $\mu_{l} \propto I$ at the metal$|$magnetic insulator interface, while  Joule heating generates a thermal gradient $\nabla T \propto I^2$.  

Spin and thermal injection trigger a spin current that might propagate to the magnetic insulating$|$detector interface. In the detector, owing to the ISHE, the impinging spin current is converted into a charge current, which  generates a voltage $V$ under open-circuit conditions. Using lock-in amplifiers, one can separate higher order  contributions in the voltage by measuring higher harmonics: 
\begin{align}
V=R_{1} I + R_{2} I^2+ ... \; \,,
\end{align}
 where $R_{i}$ is the $i$th harmonic response of the non-local resistance $R=V/I$~[\onlinecite{exptransp1}]. The first, $R_{1}$, and second, $R_{2}$, harmonic signals are a measure of,  respectively, the electrically- and thermally-generated  spin currents. An external magnetic field $\mathbf{B}$, oriented at an angle $\alpha$ within the sample plane, sets the equilibrium direction of the magnetic order parameter $\mathbf{n}$.   The effective component of the spin accumulation, $\mu_{\text{eff}}$, that exerts a spin-transfer torque on the order parameter is parallel to the magnetic order, i.e., $\mu_{\text{eff}}=\mu_{l} \cos\alpha$. Similarly, the spin current  injected via spin pumping into the detector is modulated by a factor $\cos \alpha$. Thus, the electrically-generated signal scales as $R_{1} \propto \cos^2\alpha$, while for thermal injection one has $R_{2} \propto \cos \alpha$, as shown in Figs.~\ref{SpinTransport}(b) and (c).
 
Numerous experiments have been performed using a non-local transport setup. In  this tutorial, we focus on the first experimental report  of room-temperature long-range spin transport in a magnetic insulating system. Cornelissen and coauthors~[\onlinecite{spindiff1}] investigated spin transport in YIG using the setup depicted in Fig.~\ref{SpinTransport}(a). In the data analysis, the 
spin and heat transport in the Pt$|$YIG$|$Pt heterostructure is modelled using the phenomenology presented in Sec.~IIC. 
The magnon-phonon energy relaxation length $\ell_{u}=\sqrt{\kappa/g_{uT}}$, associated with spin-preserving relaxation of magnon distribution towards the phonon temperature, is assumed to be much shorter than the magnon relaxation (i.e., spin diffusion) length $\ell_{s} \equiv \sqrt{\sigma/g_{n \mu}}$.  This can be intuitively understood as spin conserving phonon-magnon interactions can stem from Heisenberg exchange coupling, whose strength is much larger than the spin-orbit
interactions responsible for spin non-conserving processes. In this limit, by plugging Eq.~(\ref{203}) into Eq.~(\ref{210}), one obtains, in a steady-state, the spin diffusion equation 
\begin{align}
\nabla^2 \mu=\frac{\mu}{\ell_{s}}\,.
\label{FIT}
\end{align}
Fitting Eq.~(\ref{FIT}) to the experimental signal, shown in Fig.~\ref{SpinTransport}(d), leads to the estimate $\ell_{s} \sim 10  \; \mu$m, which has  been corroborated by further experiments~[\onlinecite{spindiff2},\onlinecite{spindiff3}]. It worth noting that, few years later,  the magnon-phonon energy relaxation length $\ell_{u}$  in YIG was measured to be about 250 nm~[\onlinecite{magnonrelax}, \onlinecite{magnonrelax1}], i.e., much shorter than the spin diffusion length, confirming the assumption  that underlies the analysis of the experimental data in Ref.~[\onlinecite{spindiff1}].

An alternative technique for probing  thermally-generated spin currents is offered by a longitudinal spin Seebeck  setup, comprised of a magnetic insulator sandwiched between a substrate and a metallic strip. A heat gradient applied across the sample generates a spin current in the magnetic insulator, which is then converted into a measurable voltage in the metal via ISHE. This setup, shown in Fig.~\ref{FigMP}(c), was used to detect the signatures of magnon-phonon hybridization in YIG~[\onlinecite{MP1}],  and, more recently, as a probe of thermally-driven spin transport in the easy-axis antiferromagnetic  insulator $\text{MnFe}_{2}$~[\onlinecite{SHEAF}]. 
\begin{figure}[t!]
\includegraphics[width=1\linewidth]{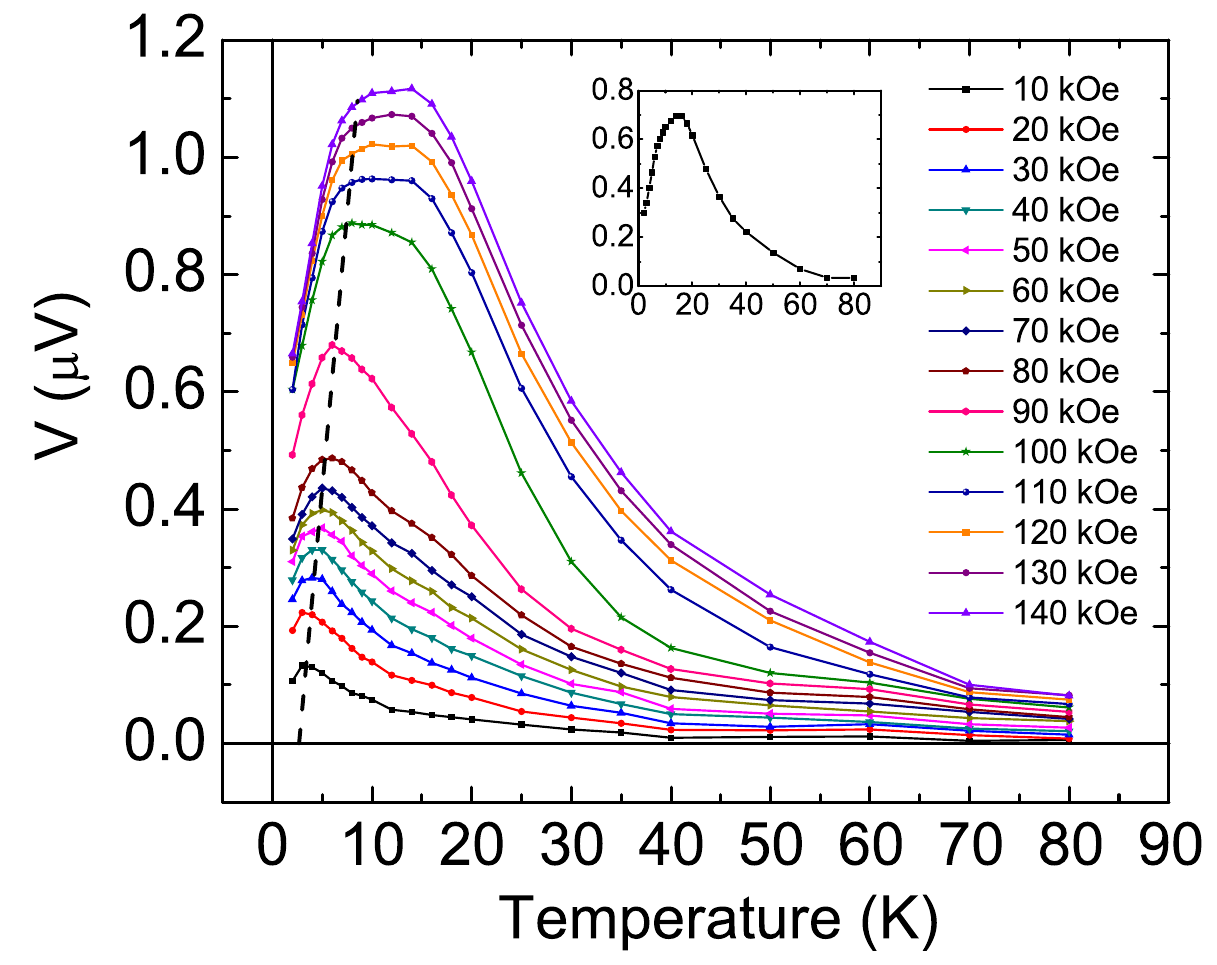}
\caption{Temperature dependence of the spin Seebeck voltage response at various magnetic fields~[\onlinecite{SHEAF}]. }
\label{FigAF}
\end{figure} 
Wu and coauthors revealed a dependence of the spin Seebeck voltage on temperature  that is not observable in ferromagnetic system. The experimental signal, shown in Fig.~\ref{FigAF}, displays peaks at temperatures much lower than the magnetic ordering temperature. The position of the peaks varies with the intensity of the applied magnetic field. While a detailed model and analysis of the data can be found in Ref.~[\onlinecite{Rezende}], here we can develop an intuitive understanding by using the concepts outlined in Sec. IIB.  $\text{MnFe}_{2}$ can be modeled as a two-sublattice uniaxial antiferromagnet, with two normal magnon modes that carry opposite spin angular momentum~(\ref{AFsz}), and, thus, generate spin currents propagating in opposite directions.  The spin current density operator $\hat{\mathbf{j}}_{s}$ reads as
\begin{align}
\hat{\mathbf{j}}_{s}=\frac{\hbar}{V} \left[ - \mathbf{v}_{ \mathbf{k}, \alpha} \hat{\alpha}^{\dagger}_{\mathbf{k}} \hat{\alpha}_{\mathbf{k}}+\mathbf{v}_{ \mathbf{k}, \beta} \hat{\beta}^{\dagger}_{\mathbf{k}} \hat{\beta}_{\mathbf{k}} \right]\,,
\label{spincurrentdensity}
\end{align} 
where $\mathbf{v}_{ \mathbf{k}, \alpha (\beta)}=\partial_{\mathbf{k}} \omega_{\mathbf{k},\alpha (\beta)}$ is the group velocity of the $\alpha$ ($\beta$) magnon mode. For an uniaxial antiferromagnet  subjected to an external magnetic field $B$  oriented along its symmetry axis,  the dispersion can be written as
\begin{align}
\omega_{\mathbf{k}, \alpha (\beta)}=\omega_{\textbf{k}} \pm \gamma B\,,
\label{322}
\end{align}
where $\omega_{\mathbf{k}}$ is the zero-field spin-wave dispersion, whose explicit form depends on the specific microscopic model.
In the absence of a magnetic field, the two magnon modes are degenerate; thus, they have equal occupation numbers in thermal equilibrium. Since they also have equal group velocities,  the spin current~(\ref{322}) vanishes. When a  magnetic field is applied,  the magnon dispersions split according to Eq.~(\ref{322}). At a given temperature, the magnon mode $\beta$ has a higher occupation density than the magnon mode $\alpha$. The difference between the occupation density of the $\alpha$ and $\beta$ modes is proportional to the splitting $2\gamma B$ between their dispersions. For temperatures lower than the bottom of the dispersion curve of the $\alpha$ mode, i.e., $T< \omega_{\mathbf{k}=0,\alpha}$, increasing the temperature increases the thermal population of the $\beta$ mode and thus the net spin current~(\ref{spincurrentdensity}). At higher temperatures, the thermal population of the $\beta$ mode start increasing, leading to a reduction of the net current~(\ref{322}).

\begin{figure*}[tbtb]
    \centering
    \includegraphics[width=0.9\linewidth]{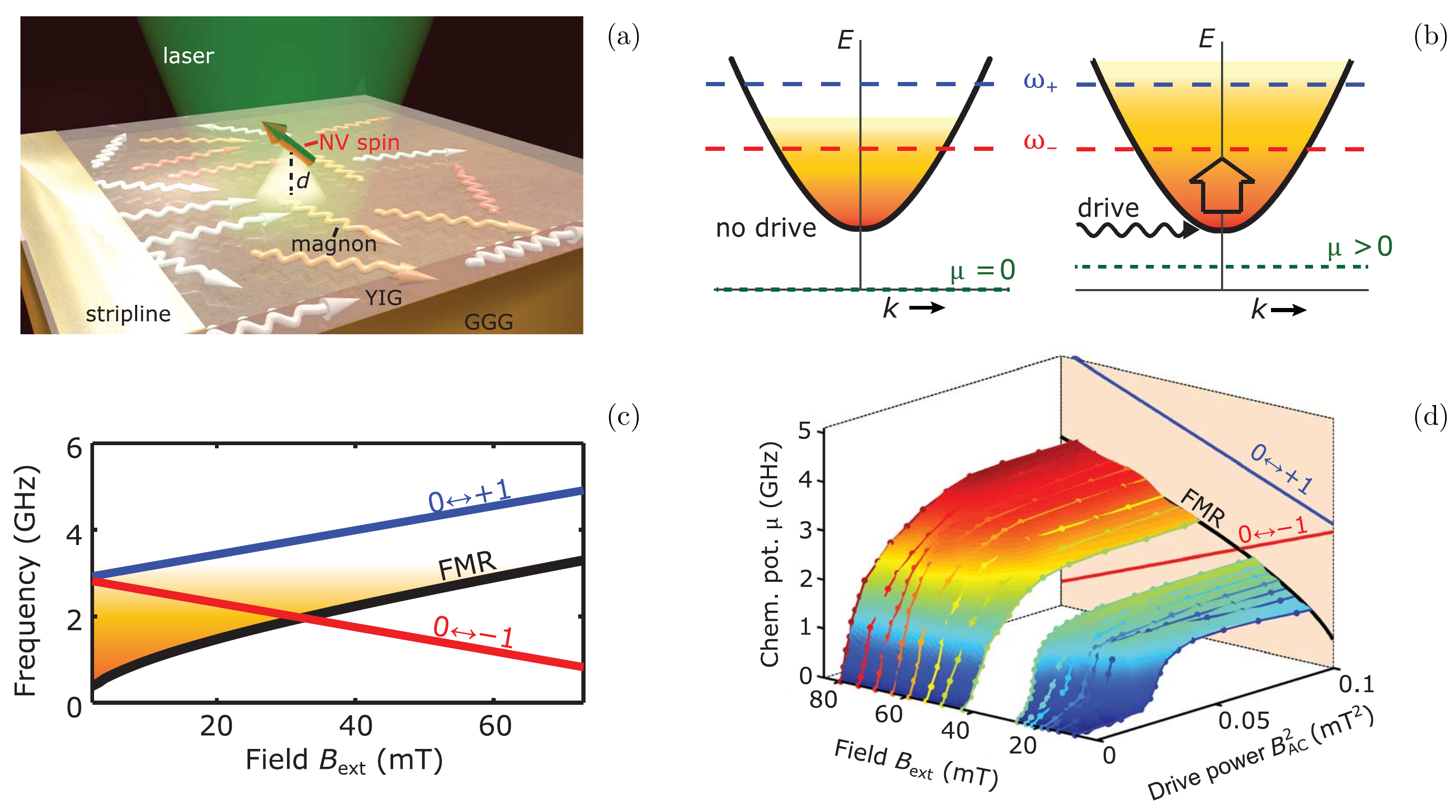}
    \caption{NV relaxometry setup and experimental data~[\onlinecite{Du2017}]. (a) A stripline excites long-wavelength magnons in a YIG film deposited on a GGG substrate. A NV center lies at a height $d$ above the sample. The magnetic noise generated by  magnons affects the NV-center relaxation rates, which are probed by a laser. (b) Dependence on the external magnetic field of the FMR (black line) and  of the NV-center relaxation frequencies  for the transitions $0 \leftrightarrow +1$ (blue line) and $0 \leftrightarrow -1$ (red line). The orange shaded area corresponds to the magnon population at room temperature. (c) In the absence of external drives, the magnon chemical potential vanishes due to dissipative interactions with the phonon bath. When a drive excites long-wavelength magnons, the thermal magnon density and, thus, the magnon chemical potential increase. (d) The increase of the magnon chemical potential is proportional to the power of the driving field $B_{AC}$, until it saturates at the ferromagnetic resonance value.}
    \label{NVDu}
\end{figure*}
Thus, the resulting temperature dependence of the spin Seebeck voltage will display a peak, as observed in the experimental data in Fig.~\ref{FigAF}. The position and the height of the peak depend on the applied magnetic field. With increasing fields, the dispersion of the $\alpha$ mode  moves  further up in energy. The thermal population of the $\beta$ mode can increase further with increasing temperatures  without populating the $\alpha$ mode, which leads to an enhanced signal at larger temperatures.

\subsection{NV-center relaxometry}

NV centers in diamond are point defects where one carbon atom in the diamond's crystal lattice is replaced by a nitrogen atom (N) and an adjacent lattice site is left empty~[\onlinecite{NV}]. Due to their exceptional sensitivity  to magnetic fields, they have recently emerged as a minimally-invasive  probe of magnetic systems, which provides a  frequency resolution not achievable by other techniques~[\onlinecite{NVreview}].
For the purpose of this tutorial, it suffices to model a NV center as a three-level spin system ($|\mathbf{S}|=1$) subjected to an external field $\mathbf{B}(\mathbf{r}_{\text{nv}})$. The external field accounts for the magnetic noise generated by an adjacent system and for a static field applied to split the degeneracy of the $m_{s}=\pm 1$ NV spin states. The NV spin $\mathbf{S}$, set at the position $\mathbf{r}_{\text{nv}}$, is oriented along its anisotropy axis $\hat{\mathbf{n}}_{\text{nv}}$, with $\hat{\mathbf{n}}_{\text{nv}} \cdot \hat{\mathbf{z}}=\cos \theta$. The corresponding Hamiltonian can be written as
\begin{align}
\mathcal{H}_{\text{nv}}=D S^2_{z} + \gamma_{e} \mathbf{S} \cdot  \mathbf{B}_{\text{nv}}(\mathbf{r}_{\text{nv}}) \,, 
\end{align}
where $D=2.87$ GHz is the ground state zero-field splitting between the $m_{s}=0$ and the degenerate $m_{s}=\pm 1$ states, and $\gamma_e= 28 \; \text{GHz} \cdot \text{T}^{-1}$  is the gyromagnetic ratio of the electronic spin. Here, we have introduced $\mathbf{B}_{\text{nv}}(\mathbf{r}_{\text{nv}})=  \mathcal{R}_{x}(\theta)\mathbf{B}(\mathbf{r}_{\text{nv}})$, where $\mathcal{R}_{x}(\theta)$ is a rotation matrix  that allows us to easily distinguish between the longitudinal and transverse components of the field with respect to the NV anisotropy axis.

NV magnetometry and relaxometry are used as probes of, respectively, static and dynamical properties of magnetic systems~[\onlinecite{NVreview}]. The quantity measured in a magnetometry experiment is the projection of the magnetic field onto the NV  anisotropy axis, $B_{\text{nv},z}$, to which the NV electron spin resonance splitting is first-order sensitive. The full vector field can be reconstructed from the field component $B_{\text{nv},z}$ and compared with the one generated by a given magnetic texture~[\onlinecite{NVreview}].

In a relaxometry setup, the relevant quantities are the field components  transverse to the NV anisotropy axis, $B_{\text{nv},x}$ and $B_{\text{nv},y}$. Up to leading order in perturbation theory, the Zeeman coupling between the NV spin and the transverse  field components  induces NV transitions between the spin states $m_{s}=0 \leftrightarrow \pm 1$ at the resonance frequency $\pm \omega$.  
For a NV-center spin set at a height $d$ above a $U$(1)- and translationally-symmetric magnetic system,   
the relaxation rate at frequency $\omega$ can be written as~[\onlinecite{FlebusQI}] 
\begin{align}
\Gamma(\omega)=& f(\theta) \int^{\infty}_{0} dk \; k^3 e^{-2kd} \left[ C_{xx}(k,\omega) +  C_{zz}(k,\omega)  \right]\,,
\label{relaxation}
\end{align}
with $f(\theta)=  (\gamma \gamma_{e})^2  ( 5 -\cos2\theta)/16 \pi$.  Here, $C_{\alpha \beta}(\mathbf{k},\omega) $ is the Fourier transform of the spin-spin correlator $C_{\alpha \beta}(\mathbf{r},\mathbf{r'}; t)=\langle\{ \hat{s}_{\alpha}(\mathbf{r}',t), \hat{s}_{\beta}(\mathbf{r},0)\} \rangle $ of the magnetic system, with $\alpha, \beta= x,y,z$.   Equation~(\ref{relaxation}) shows that the NV relaxation rate is a measure of the magnetic noise transverse, $C_{xx}$, and longitudinal, $C_{zz}$, to the equilibrium orientation of the order parameter $\mathbf{n} \parallel \hat{\mathbf{z}}$ of the magnetic system.  Invoking the HP transformations~(\ref{HP}) and~(\ref{HPAF}) for, respectively, a ferromagnetic (FM) and an antiferromagnetic (AFM) system, it is easy to see that the transverse spin-spin correlator accounts for one-magnon processes, i.e., the creation (or annihilation) of a magnon at frequency $\omega$. Thus, it is proportional to the thermal  magnon distribution function  at frequency $\omega$, i.e., 
\begin{align}
C^{\text{FM}}_{xx}(k,\omega) \propto& \; n_{\text{BE}}\left( \frac{\hbar \omega-\mu}{k_{B}T}\right)\,, \label{NVf}\\
C^{^\text{AFM}}_{xx}(k,\omega) \propto& \; n_{\text{BE}}\left( \frac{\hbar \omega  \pm \mu}{k_{B}T}\right)\,. \label{NVaf}
\end{align}

Equations~(\ref{relaxation}),~(\ref{NVf}) and~(\ref{NVaf}) show that the NV  relaxation rate can provide  a direct measurement of the magnon chemical potential, which has proven to be hardous to  perfom in spin transport setups~[\onlinecite{Cornelissen2016}]. 
The first  measurement of the magnon chemical potential via NV-center relaxometry was performed by Du and coauthors~[\onlinecite{Du2017}].  In their room-temperature setup, sketched in Fig.~\ref{NVDu}(a), a NV center is set above a YIG film.  A stripline  drives the ferromagnetic resonance of the YIG film, i.e., it excites the coherent spin dynamics obeying Eqs.~(\ref{LLG}), (\ref{132}) and (\ref{133}). The coherent spin dynamics, in turn, pumps spin angular momentum into the thermal magnon cloud. This mechanism  increases thermal magnon density and, consequently,  the magnon chemical potential~(\ref{133}), as  depicted in Fig.~\ref{NVDu}(b).  

Figure~\ref{NVDu}(c) shows the dependency of the NV-center relaxation frequencies and of the YIG ferromagnetic resonance frequency on the static external field. The transverse noise corresponding to single-magnon processes can be probed by a NV-center relaxation rate at frequencies for which the magnon thermal population is finite.
The increase in the magnon chemical potential was found to be directly proportional to the driving power, and, as predicted by Bose-Einstein statistics, to saturate at the ferromagnetic resonance frequency, as shown in Fig.~\ref{NVDu}(c).
A further analysis of the experimental data~[\onlinecite{Du2017}] lead to the extrapolation of the parameter $\eta$~(\ref{132}). 

The chemical potential of a $U$(1)-symmetric antiferromagnetic system  might be probed in a similar fashion~[\onlinecite{FlebusCP}]. The two antiferromagnetic normal modes undergo precessions with opposite handedness, as shown in Fig.~\ref{Fig1}(b); thus, they can be selectively excited by an ac field with matching polarization. Resonantly driving the coherent spin dynamics of the antiferromagnetic modes $\alpha$  and $\beta$ increases their thermal magnon populations, as dictated by Eqs.~(\ref{193}). The increase in the  thermal population of $\alpha$ ($\beta)$  leads to a increase of a negative (positive) chemical potential, which can be measured according to Eqs.~(\ref{relaxation}) and~(\ref{NVaf}). 
\begin{figure}[t!]
\includegraphics[width=1\linewidth]{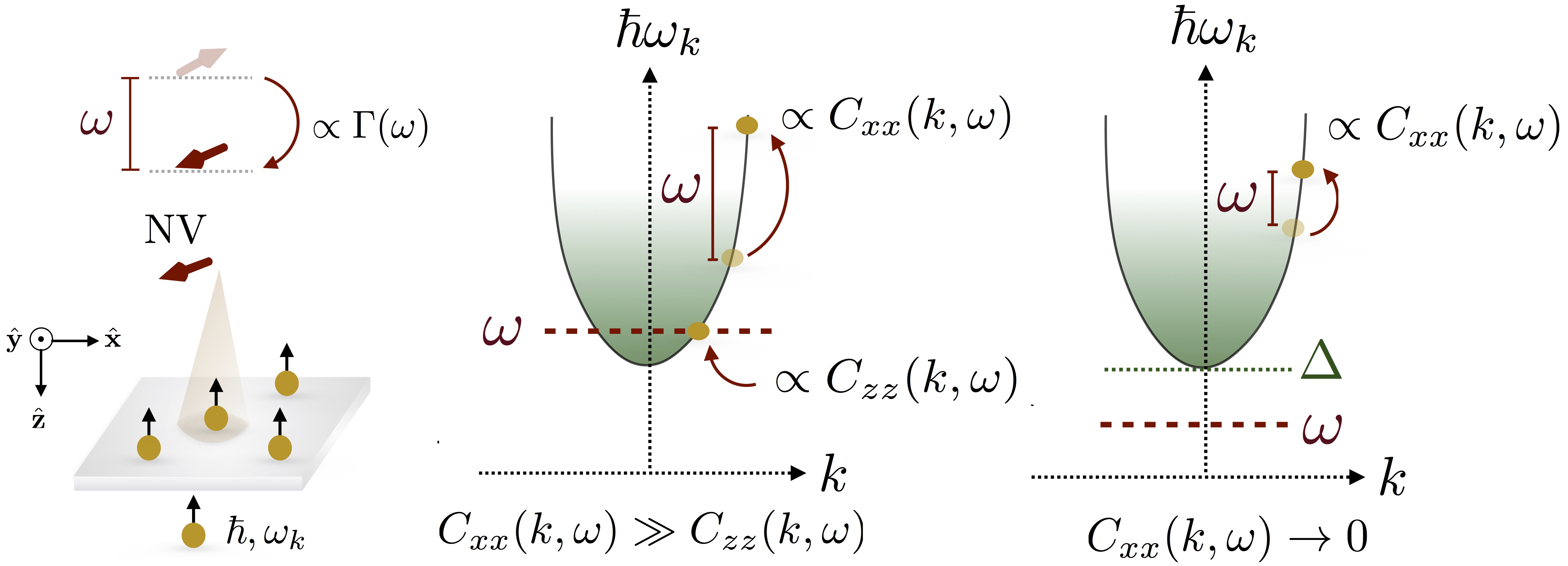}
\caption{The interaction between the NV-center spin and a nearby magnetic system, here depicted as gas of magnons with spin $\hbar$ and frequency $\omega_{k}$ (with $\omega_{k=0}=\Delta$), leads to a NV-center transition rate $\Gamma(\omega)$ with emission of energy $\omega$. When $\omega>\Delta$, the latter can result in the creation of a magnon at frequency $\omega_{k}=\omega$ or in a magnon scattering with energy gain $\hbar \omega$. These events are accounted for, respectively, by the transverse, i.e., $C_{xx}(k,\omega)$, and by the longitudinal, i.e., $C_{zz}(k,\omega)$,  spin-spin correlation function. When $\omega>\Delta$, the relaxation rate is typically dominated by one-magnon processes, i.e., $C_{xx}(k,\omega) \gg C_{zz}(k,\omega)$. Conversely, for $\omega<\Delta$, one-magnon events are suppressed, i.e., $C_{xx}(k,\omega) \rightarrow 0$~[\onlinecite{FlebusQI}].}
\label{NVQI}
\end{figure} 

In order to probe the transverse noise, the NV transition frequency $\omega$ must be larger than the  gap $\Delta$ of the spin-wave dispersion. At lower frequencies, there are no available magnon states  and, thus, Eqs.~(\ref{NVf}) and (\ref{NVaf}) vanish. For $\omega>\Delta$,  the transverse noise dominates over the longitudinal noise~[\onlinecite{Du2017},\onlinecite{FlebusQI}], while for $\omega<\Delta$ the longitudinal noise represents the leading contribution to the relaxation rate~(\ref{relaxation}), as sketched in Fig.~\ref{NVQI}.  By invoking the HP transformations~(\ref{HP}) and (\ref{HPAF}), for, respectively, a ferromagnetic and an antiferromagnetic system,  the longitudinal spin-spin correlator $C_{zz}(k,\omega)$ can be expressed in terms of two-magnon scattering processes, i.e., a magnon with frequency $\omega_{1}+\omega$  scatters into a magnon state with frequency $\omega_{1}$, or \textit{vice versa}, emitting (or absorbing) magnetic noise at frequency $\omega$. The details of how two-magnon scattering processes occur depend on the spin transport properties of the system. For diffusive magnon transport in the absence of a heat gradient~(\ref{diff}, \ref{diffAF}), the imaginary part of the spin susceptibility can be found as~[\onlinecite{FlebusQI}]
\begin{align}
\chi^{''}_{zz}(k,\omega)=\frac{\chi \hbar^2}{D}\frac{ \omega k^2}{\left(k^2+1/\ell^2_{s} \right)^2+(\omega/D)^2}\,,
\label{suscep}
\end{align}
where $\chi$ is the static uniform longitudinal susceptibility and $D=\sigma/\chi$ the spin-wave diffusion coefficient. In thermal equilibrium, the fluctuation-dissipation theorem dictates~[\onlinecite{Kubo}]
\begin{align}
C_{zz}(k,\omega)=\coth \left( \frac{\hbar \omega}{2k_{B}T} \right) \chi^{''}_{zz}\left(k,\omega\right)\,.
\label{fluctuations}
\end{align}
Wang and coauthors~[\onlinecite{Dupreparation}] recently measured the relaxation rate~(\ref{relaxation}) of a YIG film at frequencies below the magnetic gap. Using Eqs.~(\ref{suscep}) and~(\ref{fluctuations}) they extracted a spin diffusion length of $\ell_{s}\sim 1.5$ $\mu$m, which was further corroborated by a non-local spin transport measurement performed on the same sample. An analogous method can be used to probe non-invasively the spin transport properties of collinear antiferromagnets. Wang and coauthors~[\onlinecite{WangAF}] probed the time-dependent fluctuations of the longitudinal spin density of $\alpha$-$\text{Fe}_2\text{O}_3$. They estimated the spin diffusion length $\ell_{s}$ to be $3$ $\mu$m at 200 K, which is in agreement with the values reported by non-local spin transport experiments~[\onlinecite{AFdifflength}]. 

\section{Summary and perspectives}

In this tutorial, we have outlined  the fundamental properties of collective spin excitations, i.e., magnons, in collinear magnetic insulating systems. 
We have reviewed their statistical and transport properties and discussed the key ingredients of the coupling between magnetic and lattice degrees of freedom.
Making use of this  theoretical framework, we have introduced the reader to two magnon sensing techniques, i.e., spin transport setups and NV-center relaxometry.

Here we have restricted our discussion to simple monoatomic ferromagnetic and two-sublattice antiferromagnetic systems. However,  there is a variety of collinear insulators  with different crystalline structures and spin interactions that continue to be discovered and that can be addressed with the methods presented here.

Finally, while non-collinear magnetic systems do not support long-range diffusive spin transport, their spin non-conserving interactions might endow the spin-wave bands with a nontrivial topological structure.  The emergence of topologically-protected long-range propagating magnon modes and their non-Hermitian topology represent an emergent promising platform to investigate spin transport and novel magnetic phenomena.

\bibliography{aipsamp}% Produces the bibliography via BibTeX.

\end{document}